# Dual Analysis of Continuous-time Economic Dispatch and Its Price Implications

Menghan Zhang and Caisheng Wang

*Department of Electrical and Computer Engineering, Wayne State University, Detroit, MI, USA*
{mhzhang, cwang}@wayne.edu

*Abstract*--As load varies continuously over time, it is essential to provide continuous-time price signals that accurately reflect supply-demand balance. However, conventional discrete-time economic dispatch fails to capture the intra-temporal variations in load and generation. Its dual solution—the marginal price—may distort economic signals, leading to inefficient market incentives. To analyze these issues, this paper develops a continuous-time dispatch model and derives its dual formulation for price analysis. The continuous-time dispatch produces dual variables that can be interpreted as price signals. Piecewise time-indexed generation and price trajectories are then constructed through a parametric programming approach. The resulting price, represented by the Lagrange multiplier of the system-wide power balance constraint, evolves piecewise along the continuous-time load profile. Each segment corresponds to a critical region characterized by a set of active constraints. Furthermore, we discuss the impact of unit-specific ramping constraints on price implications. Results indicate that continuous-time generation and price trajectories provide deeper insights into the price distortions and inefficient incentives inherent in discrete-time formulations. The proposed methodology is validated on an illustrative 5-bus system and a modified IEEE RTS-2019.

*Index Terms*--Continuous-time dispatch, Electricity price, Generation trajectory, Parametric programming.

## I. INTRODUCTION

IN market-based power systems, the electricity spot market facilitates energy and ancillary service trading, thereby maintaining system balance and operational reliability [1]. The market clearing price can be interpreted from two perspectives. Economically, it acts as a signal of the true marginal cost, reflecting physical constraints and ensuring appropriate compensation for contributing resources [2]. Mathematically, it is a time-dependent function derived from system optimization, capable of changing at any instant of time [3]. Ideally, the dispatch aligns supply and demand, and its dual solution—the marginal price—should provide transparent and efficient incentives for both suppliers and consumers [4][5].

In practice, however, system (or market) operators adopt discrete-time formulations to ensure solvability and tractability within market time windows [6]. This discretization introduces inherent model errors, since intra-temporal variations in load and generation are not explicitly represented. With increasing renewable penetration, these short-term variations have intensified. Such fluctuations frequently activate ramping constraints, limiting the flexibility of generation scheduling. As a result, a system may have sufficient generation capacity yet lack adequate ramping capability, forcing subsequent dispatches to rely on more expensive units or reserves. While these requirements are partly driven by forecast uncertainty, they also stem from the modeling framework itself: discrete-time formulations omit explicit intra-temporal variations, leading to deterministic model errors [7].

### A. Research Motivation

To address intra-temporal variability, electricity markets have introduced ancillary services such as flexibility ramping products, fast frequency response, and regulation mileage, which enhance operational flexibility and maintain system reliability. The prices of these services are directly or indirectly linked to the dual variables of ramping constraints [8]. Although system operators do not publish an explicit "ramping price," its influence is embedded within the overall market structure through energy and ancillary service prices [2][9]. However, current market practices consider price signals only at discrete time instants, which may distort economic signals and potentially induce inefficient incentives. Therefore, without explicitly accounting for the price implications of ramping constraints, the efficiency, fairness, and transparency of market outcomes may be compromised, even with ancillary services in place.

These issues raise several key questions: What are the price implications of discrete-time modeling? At which instants along the dispatch horizon do significant price changes occur? How can short-term price variations be systematically analyzed? Moreover, price signals are published at varying time resolutions across market processes, particularly during the transition from day-ahead to real-time markets in the USA [10][11]. This paper aims to analyze these questions from a continuous-time perspective and provide a systematic framework for interpreting the price implications.

### B. Literature Review

Traditionally, spot electricity prices were defined as the Lagrange multipliers of system-wide power balance constraints [12] and were later extended to incorporate system-wide transmission constraints [13]. However, discrete-time formulations fail to capture intra-temporal variations in load and generation, leading to cases where total generation capacity is sufficient but ramping capability is inadequate. Such ramping resource shortage events cause price spikes and raise concerns about the economic rationality of the resulting prices, since ramping constraints are unit-specific rather than system-wide. The extent to which these unit-specific constraints should be reflected in price formation remains an open question.

One research stream addresses this issue by directly embedding the dual variables of ramping constraints into price formation [14]. For example, Ref. [15] shows that the rolling-horizon dispatch framework may introduce misleading market incentives due to ramping constraints. For this issue, they recommend including these dual variables so that price signals for generators with binding ramping constraints converge to their bid costs, yielding outcomes similar to a pay-as-bid scheme. However, this direct-embedding approach differentiates an otherwise identical energy commodity across market participants according to unit-specific constraints, thereby challenging the principle of uniform energy pricing.

Another approach treats ramping capability as a distinct resource—separate from energy commodities—that highlights performance differences among generating units [16]. In such frameworks, ramping costs are indirectly embedded in price formation, and the corresponding ancillary services are designed accordingly. Although these designs internalize ramping costs, the rationale for treating ramping capability as a derivative commodity remains debatable. This is because the ramping capability of generating units, similar to their generation capacity, constrains their energy production and thus impacts the marginal price of energy. As a result, the effectiveness of such designs depends on a more explicit analysis of ancillary service demand and its interaction with energy pricing.

A third research direction seeks to improve the time resolution to better approximate continuous-time results within discrete-time formulations. While higher time-resolution dispatch can reduce model approximation errors, ambiguity in defining ramping constraints persists. Finer time spans may unintentionally exacerbate price volatility, as the same ramping capability must be distributed across more time subdivisions. Moreover, increasing time resolution inevitably raises the computational burden. This may result in distorted or even ineffective market outcomes within the limited market time windows [17]. These challenges highlight the inherent limitations of discrete-time frameworks.

Notably, more accurate results require solving the dispatch problem with high time-resolution load data rather than interpolating prices or generation outputs. Some studies go beyond discrete-time formulations by modeling the dispatch results as continuous-time functions of time [18]. Theoretically, using the calculus of variations, marginal prices are derived through the Euler-Lagrange equation, thereby explicitly capturing both generation trajectories and ramping rates [19]. Nevertheless, the integration of such continuous-time formulations into practical market mechanisms remains limited. The approximations employed in these methods do not fully trace continuous-time load profiles. Since the price signals are highly sensitive to variations in input data, these approximations may still introduce distortions.

In summary, modern power systems employ different pricing schemes for energy and ancillary services within discrete-time frameworks. Ancillary services related to ramping capabilities are critical for maintaining operational reliability and require fair and transparent market mechanisms for their procurement and pricing. Failure to explicitly quantify the price implications of ramping constraints may compromise the transparency and fairness of both energy and ancillary service market outcomes. Therefore, it is essential to re-examine price formations and price signals from a continuous-time perspective and investigate the price implications of ramping constraints to ensure transparency, fairness, and economic efficiency.

*C. Contributions*

In our previous work [20], we developed a methodology to build generation trajectories that balance continuous-time load profiles, focusing on the solvability of the variational problem and the satisfaction of infinite-dimensional constraints. However, the dual analysis was not performed, and the price implications remained unexplored.

In this paper, we address this gap by deriving the dual solution of the continuous-time dispatch. Starting from the primal problem and its Lagrange multipliers, we demonstrate how continuous-time generation trajectories give rise to marginal prices. These prices correspond to the Lagrange multipliers of the continuous-time power balance constraint and evolve piecewise along the continuous-time load profiles, with each segment corresponding to a distinct critical region defined by a set of active constraints.

A distinctive feature of our approach is the use of parametric programming to partition the continuous-time load profile into polytope-based critical regions. Each critical region produces corresponding continuous-time generation and price trajectories. By introducing the concept of critical regions, the parameter space is partitioned according to the set of active constraints, enabling the identification of time instants at which ramping constraints become binding. This analysis reveals the sensitivity of prices to constraint activation in the primal-dual problem, where prices may exhibit discontinuities.

Furthermore, the continuous-time results reveal that unit-specific ramping constraints, as reflected in generation trajectories, determine the precise instants at which price changes occur. The continuous-time perspective emphasizes the coupling effects of temporal constraints across the dispatch horizon and, therefore, provides a more comprehensive interpretation of price distortions and inefficient incentives induced by discrete-time formulations. The proposed methodology is validated on an illustrative 5-bus system and a modified IEEE RTS-2019.

*D. Term and Notation*

The following terms and notations are used throughout the paper. The *period*, denoted as $\mathcal{T}=[S,T]$, represents the overall time span of the dispatch horizon, with $t_0=S$ as the start and $t_N=T$ as the end. The *interval* refers to the subdivision of the overall period, such that $\mathcal{T}$ is subdivided into $N$ intervals. Each interval is denoted as $\mathcal{T}_n=[t_n, t_{n+1}]$, $\mathcal{T}=\cup_{n=0}^{N-1}\mathcal{T}_n$, where $n$ is the interval index. The time length of $\mathcal{T}_n$ is $\Delta t_n$, where $\Delta t_n=t_{n+1}-t_n$. The *range* refers to any subperiod covering multiple intervals but shorter than $\mathcal{T}$. A *segment* refers to the continuous-time load profile within an interval $\mathcal{T}_n$. The continuous-time load profile is denoted as $D(t)$, and the sampled discrete-time loads are represented by $D(t_n)$. The *interval endpoints* $t_n$ and $t_{n+1}$ correspond to the *segment endpoints* $D(t_n)$ and $D(t_{n+1})$ at those instants. The generation and price trajectories are expressed as piecewise affine functions, each defined over a segment of the continuous-time load profile.

## II. THEORETICAL PRIMAL-DUAL PROBLEM FORMULATION

We first formulate the primal economic dispatch problem, derive the corresponding dual formulation, and then discuss the assumptions and simplifications considered in this paper.

## A. Primal Economic Dispatch Problem

The objective of economic dispatch is to determine the cost-minimizing unit outputs that satisfy the power balance over the period $\mathcal{T}$. In the discrete-time formulation, the period $\mathcal{T}$ is divided into $N$ intervals and each interval $\mathcal{T}_n$ has a time length $\Delta t_n$. The decision variables $\mathbf{G}(t_n)$ of generation are optimized to match $D(t_n)$, leading to the following discrete-time dispatch problem.

$$\min_{\mathbf{G}(t_n)} \sum_{n=0}^{N-1} \big( C(\mathbf{G}(t_n)) \big) \Delta t_n \tag{1}$$

$$\text{s.t.} \quad \sum_K G_k(t_n) = D(t_n), \forall n \tag{2}$$

$$G_k(t_n) \in \mathcal{G}_k, \forall k, n \tag{3}$$

$$\mathcal{G}_k = \big\{ G_k(t_n) \,|\, G_k^{\min} \le G_k(t_n) \le G_k^{\max};$$
$$R_k^{\text{down}} \Delta t_n \le \big(G_k(t_{n+1}) - G_k(t_n)\big) \le R_k^{\text{up}} \Delta t_n \big\} \tag{4}$$

where $k$ is the unit index and $K$ is the total number of units. The expression $C(\mathbf{G}(t_n))=\Sigma_K C_k(G_k(t_n))$ denotes the sum of the production costs for all units at time $t_n$. $\mathcal{G}_k$ represents the operational limitation set of individual unit $k$. The superscripts min/max are the minimum/maximum generation output of unit $k$. $R_k^{\text{down}}/R_k^{\text{up}}$ are the downward/upward ramping capabilities of unit $k$ (in MW per time unit).

The objective function (1) minimizes the total production cost with respect to $\mathbf{G}(t_n)=(G_1(t_n),\ldots,G_K(t_n))^\top$ over all intervals. Constraints (2) enforce the system-wide power balance at each segment endpoint. Constraints (3)-(4) impose operational limitations (unit-specific constraints) on each unit $k$, including generation capacity constraints and ramping constraints.

Typically, unit ramping is approximated by a linear trajectory within each interval, represented by the finite difference of generation outputs between two dispatch points. As the time resolution increases ($\Delta t_n \to 0$), the discrete-time formulation converges to the continuous-time formulation. In the limit of infinitely fine subdivisions of the period $\mathcal{T}$ ($\Delta t_n \to 0, N \to \infty$), the generation derivative function of unit $k$ is given in (5).

$$G_k'(t) \triangleq \lim_{\Delta t_n \to 0} \frac{\big(G_k(t_{n+1}) - G_k(t_n)\big)}{\Delta t_n}, \forall t \in \mathcal{T}_n \tag{5}$$

Building on (5), we extend the formulation to continuous-time dispatch, where finite sums are replaced by integrals. The objective is to build generation trajectories $\mathbf{G}(t)$ that balance the continuous-time load profile $D(t)$. Following Refs. [7][18][20], the primal problem of continuous-time dispatch is formulated as follows.

$$\min_{\mathbf{G}(t)} \int_{\mathcal{T}} \big( C(\mathbf{G}(t)) \big) dt \tag{6}$$

$$\text{s.t.} \quad \sum_K G_k(t) = D(t), \forall t \tag{7}$$

$$G_k(t) \in \mathcal{G}_k, \forall k, t \tag{8}$$

$$\mathcal{G}_k = \big\{ G_k(t) \,|\, G_k^{\min} \le G_k(t) \le G_k^{\max};$$
$$R_k^{\text{down}} \le G_k'(t) \le R_k^{\text{up}} \big\} \tag{9}$$

where $C(\mathbf{G}(t))=\Sigma_K C_k(G_k(t))$ represents the sum of production costs of all units at time $t$.

The objective function (6) minimizes the total production cost with respect to $\mathbf{G}(t)=(G_1(t),\ldots,G_K(t))^\top$ throughout $\mathcal{T}$. Unlike the discrete-time formulation in (1)-(4), the constraints in model (6)-(9) are expressed as infinite-dimensional and differential constraints. The operational limitation set $\mathcal{G}_k$ is modified to represent the continuous-time ramping constraints.

## B. Dual Formulation and Locational Marginal Prices

The key differences between the discrete-time and continuous-time dispatch models lie in the use of finite sums versus integrals in the objective function, as well as in the representation of constraints. Both formulations can be expressed in the following compact form.

$$\min_x \mathcal{J} = \langle f, x \rangle \quad \text{s.t.} \quad \lambda : A_{\text{eq}} x = b_{\text{eq}}, \mu : A_{\text{ie}} x \le b_{\text{ie}}. \tag{10}$$

where $\mathcal{J}$ denotes the total production cost. $f$ denotes the vector of marginal cost coefficients. $x$ denotes the primal decision variables. The inner product $\langle \cdot, \cdot \rangle$ is defined as a finite sum in the discrete-time case and as an integral in the continuous-time case. $A_{\text{eq}}/A_{\text{ie}}$ are the coefficient matrices of equality/inequality constraints. $b_{\text{eq}}/b_{\text{ie}}$ are the constant vectors. $\lambda/\mu$ are the Lagrange multipliers associated with the equality/inequality constraints.

In this paper, only the system-wide power balance constraint is explicitly considered. Hence, equality constraints correspond to system-wide power balance constraints, whereas inequality constraints correspond to unit-specific constraints. Specifically, $b_{\text{eq}}$ denotes the load profile, represented as $D(t)$ in the continuous-time case or $D(t_n)$ in the discrete-time case. The ramping constraints (4) and (9) associated with $A_{\text{ie}}$ involve a differential operator in the continuous-time case and a finite-difference matrix in the discrete-time case.

As $\Delta t_n \to 0$, the continuous-time formulation in (6)-(9) can be regarded as the limit of the discrete-time formulation. It can be interpreted as a generalization of the vector inner product to an infinite-dimensional decision space. Differential constraints remain linear because differentiation is a linear operator in the limit of infinitely fine subdivisions. Accordingly, the Lagrangian of model (10), with load profiles $D(t)$ or $D(t_n)$, is given by (11).

$$\mathcal{L}(b_{\text{eq}}) = \mathcal{L}(D(t)) \text{ or } \mathcal{L}(D(t_n)) = \mathcal{L}(x, \lambda, \mu)$$
$$= \mathcal{J} + \langle \lambda, b_{\text{eq}} - A_{\text{eq}} x \rangle + \langle \mu, b_{\text{ie}} - A_{\text{ie}} x \rangle \tag{11}$$

By minimizing the Lagrangian with respect to the primal variables $x$, the dual problem can be expressed as follows.

$$\max_{\lambda, \mu} \langle \lambda, b_{\text{eq}} \rangle + \langle \mu, b_{\text{ie}} \rangle \quad \text{s.t.} \quad A_{\text{eq}}^{\text{ad}} \lambda + A_{\text{ie}}^{\text{ad}} \mu = f, \mu \ge 0 \tag{12}$$

where the superscript ad denotes the adjoint operator. In the discrete-time case, the adjoint operator is the transpose under the Euclidean inner product. In contrast, in the continuous-time case, it is defined with respect to the inner product of function spaces and involves integration by parts.

The dual problem (12) and its associated variables $\lambda/\mu$ have well-established economic interpretations. In electricity spot markets, locational marginal prices (LMPs) are obtained from the dual variables of the economic dispatch problem. We use the LMP as the market clearing price for the subsequent market analysis. The system-wide Lagrange multiplier $\lambda$, associated with the power balance constraints, represents the LMP. Meanwhile, the unit-specific Lagrange multipliers $\mu$, corresponding to the operational limitations $\mathcal{G}_k$, represent the shadow prices of unit-specific constraints.

We now analyze the continuous-time price signal from (10)-(12). The Lagrange multiplier $\lambda$ can be interpreted as the subgradient of the optimal cost with respect to the load profiles. Formally, let $\mathcal{J}^*$ denote the optimal cost function associated with a given load profile $D(\tau)$. A small incremental load perturbation $\delta D(\tau)$ over the interval $[\tau, \tau+\delta t]$ induces a change $\delta \mathcal{J}^*$ in the optimal cost as expressed in (13).

$$\delta \mathcal{J}^* = \langle \lambda^*, \delta D(\tau) \rangle + \mathbf{O}(\|\delta D(\tau)\|) \quad (13)$$

where the superscript * denotes the optimal results $(x^*, \lambda^*, \mu^*)$ corresponding to the given load profile $D(\tau)$, and $\mathbf{O}(\|\delta D(\tau)\|)$ represents higher-order terms that vanish faster than $\|\delta D(\tau)\|$. As $\delta D(\tau) \to 0$, these higher-order terms can be neglected, and $\delta \mathcal{J}^*$ quantifies the incremental system cost due to the infinitesimal load perturbation $\delta D(\tau)$.

$$\delta \mathcal{J}^* \approx \langle \lambda^*, \delta D(\tau) \rangle \quad (14)$$

Consider a single interval. As $\delta t \to 0$, (14) takes the form (15).

$$\delta \mathcal{J}^* = \sum_{n=0}^{N-1} \lambda^*(t_n) \delta D(\tau) \Delta t_n, t_n = \tau, \Delta t_n = \delta t, N = 1 \quad (15)$$

In the continuous-time case ($N \to \infty$), (14) takes the form (16).

$$\delta \mathcal{J}^* = \int_\tau^{\tau+\delta t} \lambda^*(t) \delta D(t) dt \approx \sum_{n=0}^{N-1} \lambda^*(t_n) \delta D(\tau) \Delta t_n \quad (16)$$

When the interval $[\tau, \tau+\delta t]$ is extended to the full period $\mathcal{T}$, the discrete-time case corresponds to increasing $N$, whereas the continuous-time case corresponds to extending the integration domain. In the remainder of this paper, we explicitly use $t$ (continuous-time) or $t_n$ (discrete-time) to distinguish the two formulations.

As shown in (13)-(16), the discrete-time and continuous-time formulations coincide in the limit. $\lambda(t)$ and $\lambda(t_n)$ quantify the marginal cost of serving an incremental load, measured in dollars per MW per unit of time. As shown later, $\lambda(t)$ represents the price trajectories (LMP) and serves as the foundation for the subsequent dual analysis.

*C. Assumptions and Simplifications*

We assume that the continuous-time load profile $D(t)$ is known in advance and represented as a single-valued, differentiable function over the full period $\mathcal{T}$. This paper focuses on the dual solution of the continuous-time dispatch and does not address nonconvexities arising from unit commitment. Commitment results are assumed to be known and fixed throughout $\mathcal{T}$. Accordingly, pricing schemes such as convex-hull pricing, which primarily address integer-related nonconvexities, are beyond the scope of this paper. For clarity, startup, shutdown, and no-load costs—typically associated with unit commitment—are omitted in production costs, since commitment results are predetermined.

Another simplification is the omission of system-wide constraints such as transmission limits. Although these constraints influence price and may constitute explicit price components in practice, they primarily affect spatial rather than temporal variations and are therefore omitted here. This simplification allows the methodology to focus on continuous-time ramping constraints without introducing additional spatial complexity. If transmission constraints were included, additional Lagrange multipliers $\mu(t)$ or $\mu(t_n)$ would appear. In that case, the inequality constraints would need to be subdivided further to distinguish system-wide constraints, whereas model (10) here only captures unit-specific limitations.

### III. DUAL RESULTS OF ITERATIVE DISPATCH METHODOLOGY

The continuous-time economic dispatch problem constitutes a variational optimization problem with differential constraints, which is generally intractable to solve directly in its original form. In our earlier work [20], a parametric programming approach was used to iteratively build feasible generation trajectories. We now extend the methodology to derive parametric solutions for the associated dual variables.

*A. Extension of the Iterative Dispatch Methodology*

The iterative dispatch methodology in [20] consisted of two stages—trajectory construction and constraint verification—designed to build feasible trajectories of primal variables. The framework is here extended to derive continuous-time dual trajectories by introducing corresponding dual formulations and performing detailed dual analysis, as outlined in Fig. 1.

In this paper, we consider an hourly period $\mathcal{T}=[S,T]=[0,60]$ minutes. This setup reflects current day-ahead market practices, where dispatch is scheduled at an hourly resolution and refined later in the real-time market. Before the iteration begins, the procedure is initialized with input data, including the hourly unit commitments, hourly economic dispatch, and the continuous-time load profile $D(t)$. In each iteration, the updating range is determined based on the feasibility of the generation and price trajectories obtained in the previous iterations. The boundary values of the primal decision variables within the updating range are specified from the discrete-time adaptive model in the last constraint verification stage. During the iterative dispatch, both primal (generation) and dual (price) trajectories are updated and saved until the convergence criterion is satisfied. If feasibility cannot be maintained, the iteration terminates with an infeasible outcome.

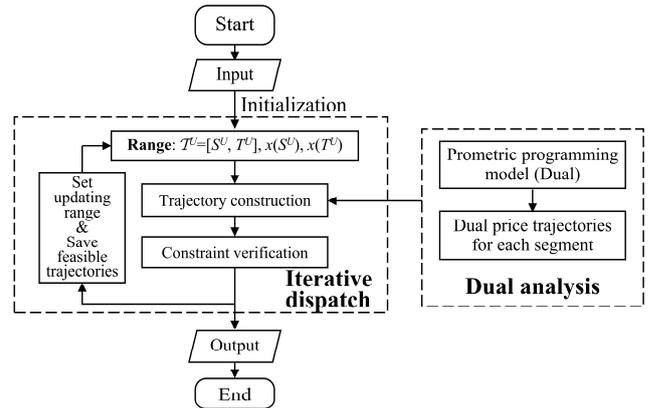

Fig. 1. Overview of the iterative dispatch methodology and its extension for dual analysis.

The main focus of this paper lies in the dual analysis, which requires clarification of the relationship between the primal and dual formulations. The continuous-time formulations can be written as an infinite-dimensional optimization, from which the Karush-Kuhn-Tucker (KKT) conditions and time-dependent dual variables can be formally derived. However, the direct use of these infinite-dimensional KKT conditions is impractical due to the functional nature of the dual variables, the presence of differential constraints, and the fact that the primal problem is solved through an iterative dispatch methodology rather than as a single closed-form optimization problem. As a result, the dual variables and price trajectories are not readily available from the primal solution alone. To facilitate a tractable analysis, the dual problem is constructed through a parametric representation of the primal model. The resulting unified primal-dual formulation recasts the infinite-dimensional problem into a finite-dimensional parametric form, from which both generation and price trajectories can be systematically derived.

*B. Primal-dual Model for Trajectory Construction*

In the trajectory construction stage, the continuous-time load profile is segmented according to changes in the active constraint sets. For each segment, generation and price trajectories are derived using a primal-dual parametric programming model, with time $t$ serving as the parameter to capture the evolution of decisions. Since the load profile $D(t)$ is explicitly time-dependent and cannot be represented within the linear formulation, we introduce a distinct parameter $D^P$ to capture the variations in $D(t)$. The primal-dual problem can be expressed as a two-dimensional parametric programming model in (17) within the updating range $\mathcal{T}^U=[S^U, T^U]$.

(Primal) $\quad \min_{x(\theta)} f^\top x(\theta)$

s.t. $\quad \lambda(\theta): A_{eq} x(\theta) = b_{eq} + E_{eq}\theta$

$\quad \mu(\theta): A_{ie} x(\theta) \leq b_{ie} + E_{ie}\theta$

(Dual) $\quad \max_{\lambda(\theta),\mu(\theta)} \left(b_{eq}+E_{eq}\theta\right)^\top \lambda(\theta) + \left(b_{ie}+E_{ie}\theta\right)^\top \mu(\theta)$

s.t. $\quad A_{eq}^\top \lambda(\theta) + A_{ie}^\top \mu(\theta) = f, \mu(\theta) \geq 0$ (17)

where the primal-dual decision variables $x(\theta)$, $\lambda(\theta)$ and $\mu(\theta)$ depend on parameter $\theta=[t, D^P]^\top$ with $t \in \mathcal{T}^U$. Here, $\mathcal{T}^U$ denotes the updating range determined either at initialization or during the subsequent constraint verification stage. $E_{eq}/E_{ie}$ are the coefficient matrices of $\theta$. Here, $b_{eq}$ is set to zero, and thus, the $D^P$ fully describes the $D(t)$.

The explicit formulation of model (17) follows our earlier work [20] with modifications to the ramping constraints relative to model (1)-(4). In particular, the primal decision variables $x(\theta)$ must match the given dispatch results $x(S^U)$ and $x(T^U)$ at the start $t=S^U$ and the end $t=T^U$, respectively.

The model (17) is a linear, finite-dimensional problem that treats time and load profiles as two distinct parameters. It can be interpreted as a parametric formulation at any instant $t$, with integration and summation terms omitted. The dispatch results of unit $k$ are specified at the start and end of $\mathcal{T}^U$, while ramping constraints are represented by linear approximations. Although this approximation may affect multipliers of unit-specific constraints, it does not essentially alter those of the system-wide equality constraint. Hence, the definition of the LMP remains unchanged, as it is determined by the Lagrange multipliers of the system-wide balance constraints.

The objective of the primal-dual model is to identify feasible decision variables that share an identical set of active constraints and to represent the corresponding primal-dual solutions as piecewise affine functions. Solving the model in (17) yields polyhedral partitions in the parameter space $\theta$. Once the duality conditions are satisfied, the continuous-time load profiles $D(t)$ can be partitioned into multiple segments corresponding to the associated critical regions.

Let $\mathcal{C}(\theta)$ denote the set of parametric inequality constraints of the primal model in (17). The superscript $P$ denotes the parametric solution obtained from (17). For a given $\theta$, the inequality constraints can be classified into active and inactive sets by evaluating $\mathcal{C}(\theta)$ at the solution $x^P(\theta)$.

$$\begin{cases} \mathcal{N}_i(\theta) \triangleq \left\{ j \in \{1,\cdots,m\} \,\middle|\, A_{ie,j} x^P(\theta) - b_{ie,j} - E_{ie,j}\theta < 0 \right\} \\ \mathcal{A}_i(\theta) \triangleq \left\{ j \in \{1,\cdots,m\} \,\middle|\, A_{ie,j} x^P(\theta) - b_{ie,j} - E_{ie,j}\theta = 0 \right\} \end{cases} \quad (18)$$

where $\mathcal{N}_i(\theta)$ and $\mathcal{A}_i(\theta)$ represent the inactive and active constraint sets, with $\mathcal{C}(\theta)=\mathcal{N}_i(\theta) \cup \mathcal{A}_i(\theta)$. Here, $i$ denotes the index of the critical region.

The critical region $H_i$ is defined by the inactive constraint set $\mathcal{N}_i(\theta)$, KKT conditions and parameter bounds. Within each $H_i$, $\mathcal{N}_i(\theta)$ and $\mathcal{A}_i(\theta)$ remain unchanged, and the associated primal and dual variables can be expressed as affine functions of $\theta$.

$$\begin{bmatrix} x_i^P(\theta) \\ \lambda_i^P(\theta) \end{bmatrix} = \begin{bmatrix} A_i^x \\ A_i^\lambda \end{bmatrix} \theta + \begin{bmatrix} B_i^x \\ B_i^\lambda \end{bmatrix} = \begin{bmatrix} A_i^x \\ A_i^\lambda \end{bmatrix} \begin{bmatrix} t \\ D^P \end{bmatrix} + \begin{bmatrix} B_i^x \\ B_i^\lambda \end{bmatrix}, \theta \in H_i \quad (19)$$

where $A_i^x$, $B_i^x$, $A_i^\lambda$ and $B_i^\lambda$ denote constant coefficient matrices/vectors associated with affine functions in the critical region $H_i$.

Each interval $\mathcal{T}_n=[t_n, t_{n+1}]$ corresponds to a distinct segment of $D(t)$. If the dispatch problem is feasible, each segment is contained within the corresponding $H_i$. By introducing $D(t)$ in place of $D^P$, the trajectories of the primal and dual variables can then be explicitly expressed.

$$\begin{bmatrix} x_n^C(t) \\ \lambda_n^C(t) \end{bmatrix} = \begin{bmatrix} A_i^x \\ A_i^\lambda \end{bmatrix} \begin{bmatrix} t \\ D(t) \end{bmatrix} + \begin{bmatrix} B_i^x \\ B_i^\lambda \end{bmatrix}, t \in \mathcal{T}_n \quad (20)$$

where the region index $i$ is associated with the interval index $n$, indicating that the load profiles $D(t)$ in $\mathcal{T}_n$ lie within the corresponding critical region $H_i$. The superscript $C$ denotes trajectories obtained from the trajectory construction stage.

In each iteration, the feasibility of the continuous-time dispatch problem is verified. This verification ensures that the continuous-time load profiles $D(t)$, for $t \in \mathcal{T}$, lie entirely within the corresponding critical regions. If this condition is not satisfied, the iteration terminates and the dispatch problem is deemed infeasible. Otherwise, the continuous-time generation trajectories $x^C(t)=[x_0^C(t), \ldots, x_{N-1}^C(t)]$ and price trajectories $\lambda^C(t)= [\lambda_0^C(t), \ldots, \lambda_{N-1}^C(t)]$ are obtained by gathering the results across all intervals $\mathcal{T}_n$. These trajectories are then passed to the constraint verification stage for further feasibility checking.

*C. Generation and Price Results in Constraint Verification*

The core task of the constraint verification stage is to ensure that the generation trajectories satisfy continuous-time ramping constraints and remain continuous. With the dual analysis included, it is also necessary to verify that the obtained continuous-time price trajectories (signals) are consistent with the generation trajectories. Notably, the linear model (17) enforces ramping constraints only at interval endpoints; therefore, trajectory continuity must be explicitly verified with respect to the differential constraints.

In this stage, the discrete-time adaptive model, previously developed in [20], is employed to facilitate the verification of continuous-time ramping constraints. The period $\mathcal{T}$ is divided into $N$ adaptive intervals $\mathcal{T}_n$, and the corresponding discrete-time adaptive dispatch is formulated in (21). The decision variables $x^D(t_n)$ are used to verify the continuity of generation trajectories at interval endpoints.

$$\min_{x^D(t_n)} \langle f, x^D(t_n) \rangle = \sum_{n=0}^{N-1} f^\top x^D(t_n) \Delta t_n$$
$$\text{s.t.} \quad A_{eq} x^D(t_n) = b_{eq}$$
$$A_{ie} x^D(t_n) \leq b_{ie} \quad (21)$$

where the superscript $D$ denotes the solution obtained from the discrete-time adaptive dispatch model (21).

Physically, generation trajectories are inherently continuous and unlikely to exhibit sudden jumps. Accordingly, the continuity of $x_n^C(t)$ at non-endpoints is verified through the differential constraints in $\mathcal{G}_k$. The derivatives of the primal decision variables $x_n^C(t)$ must remain within the ramping capability bounds $R_k^{\text{down}}/R_k^{\text{up}}$; otherwise, ramping violations are detected and new interval endpoints are identified. At the interval endpoints, continuity is verified by comparing the affine function values of adjacent segments, as given in (22).

$$\lim_{\varepsilon \to 0^+} x_{n-1}^C(t_n - \varepsilon) = x^D(t_n) = \lim_{\varepsilon \to 0^+} x_n^C(t_n + \varepsilon) \quad (22)$$

The mismatch error at the interval endpoints serves as the convergence criterion, as defined by (23). Continuity at an interval endpoint $t_n$ is deemed satisfied only when the mismatch error is below the predefined tolerance.

$$\begin{cases} \Delta_n(t_n) = x^D(t_n) - x_n^C(t_n) \\ \Delta_{n-1}(t_n) = x^D(t_n) - x_{n-1}^C(t_n) \end{cases} \quad (23)$$

Following [20], the above verification process examines the differential constraints, identifies feasible trajectories, and determines the updating range for the next iteration. However, this verification is not directly applicable to the dual variables. The price signal reflects the bids of the marginal unit and may exhibit abrupt changes when the marginal unit switches. Notably, since the price signals are derived from the primal formulation, they can be analyzed jointly with the verified generation trajectories, ensuring consistency in interpretation.

Building on the demand-side analysis in Section II.B, this section examines price implications from the supply perspective. Without considering transmission constraints in this paper, the entire system is characterized by a single uniform price at any instant $t$. As shown in Fig. 2, an incremental load (infinitesimal load perturbation) $\delta D(t)$ must be balanced by the marginal unit, resulting in a generation adjustment $\delta x_k(t)=\delta D(t)$. The marginal cost associated with this incremental output $\delta x_k(t)$ of the marginal unit defines the price $\lambda(t)$ at that instant. Accordingly, the continuous-time price trajectory $\lambda(t)$ represents the continuous extension of these instantaneous marginal costs over the entire period $\mathcal{T}$. From this joint supply-demand perspective, the price is defined by the system-wide balance constraint but indirectly shaped by unit-specific operational limitations through switches in the marginal unit.

Notably, $x^C(t)$ and $\lambda^C(t)$ have already been obtained in the trajectory construction stage, where the critical region $H_i$ and its associated sets $\mathcal{N}_i(\theta)$ and $\mathcal{A}_i(\theta)$ are also known. These results enable the identification of the marginal unit at any instant $t$ based on operational limitations. For example, units operating at their maximum or minimum output levels within the active constraint set are first excluded. Among the remaining units, the one that provides $\delta x_k(t)$ at the lowest marginal cost is denoted as $f_k$. If the price trajectories satisfy $\lambda^C(t)=f_k$ at that instant, the generation and price trajectories are considered consistent within the corresponding critical region $H_i$.

In summary, the Lagrange multiplier $\lambda^C(t)$ at any instant $t$ must correspond to the marginal unit in the current generation trajectory; hence, the price trajectories are regarded as valid. Furthermore, price changes are considered valid only when they coincide with switches in the marginal unit; otherwise, they are deemed invalid, and a new interval endpoint must be introduced at that instant in the next iteration.

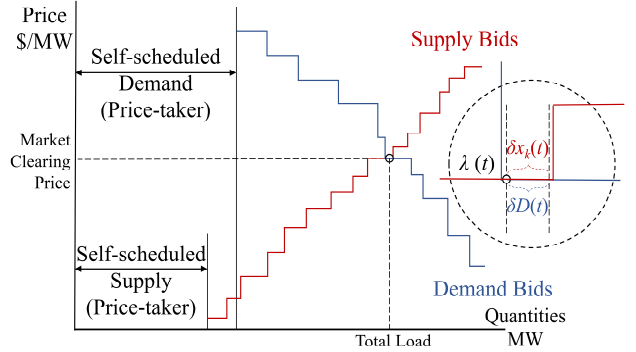

Fig. 2. Illustration of an infinitesimal load perturbation, the corresponding incremental output of the marginal unit, and their implications for the LMP.

### D. Price Discontinuities and Methodological Suboptimality

From the above analysis, the price signals can be interpreted as shadow prices associated with the supply-demand balance. However, price signals at the interval endpoints have not yet been discussed. Price signals naturally exhibit discontinuities at these points due to switches in the marginal unit. These price jumps should not be viewed as distortions but as natural outcomes. Moreover, since the marginal cost functions of units are typically piecewise linear, they yield piecewise price trajectories. The price trajectory $\lambda(t)$ evolves continuously within each active constraint set but may exhibit jumps at instants when the marginal unit switches and an inactive constraint becomes binding.

Regarding the generation trajectories, the switching of marginal units can occur smoothly as one unit gradually takes over output from another. By contrast, prices are determined by discrete unit bids, and the transition of marginal units inevitably produces jumps. In practice, such price jumps are economically rational: when the marginal unit switches, a unit-specific constraint becomes binding, leading to a shift in the marginal cost from the previous marginal unit to the new one. From the critical-region perspective, this indicates that as the load profile transitions from one critical region to another, a previously inactive constraint becomes binding, and the previous marginal unit reaches its limit, causing a switch to the new one.

If the marginal unit switches from unit $k$ to unit $k+1$, two prices naturally coexist at that instant:

$$\lim_{\varepsilon \to 0^+} \lambda_{n-1}^C(t_n - \varepsilon) = f_k, \quad \lim_{\varepsilon \to 0^+} \lambda_n^C(t_n + \varepsilon) = f_{k+1} \quad (24)$$

From a discrete-time perspective, price jumps may introduce price differences within the affected interval and cause inefficient incentives. Within a discrete-time framework, a price jump cannot be timely detected, thereby distorting the price signal until the next dispatch point. When such a price is used for settlement, it may create inefficient incentives. Nevertheless, from a continuous-time perspective, the change occurs instantaneously $\delta t \rightarrow 0$, and its impact is negligible. This implies that, once the switching instant is identified, the error induced by price jumps can be bounded to a manageable level. Thus, price jumps are inherent features of continuous-time dispatch rather than distortions of economic signals. To avoid ambiguity arising from the coexistence of two price values, this paper adopts the left-hand-side limit as the price at that instant.

In addition to price discontinuities, another intrinsic property of this methodology is suboptimality. Ramping constraints are imposed at fixed boundary points, and intertemporal behavior is linearly approximated via a multi-parametric programming formulation. Although this formulation satisfies the optimality conditions of the primal-dual problem, it does not guarantee the global optimum of the full continuous-time formulation. The resulting trajectories are therefore suboptimal. Notably, this suboptimality represents a practical trade-off between solvability and accuracy while retaining clear economic interpretations. To rigorously prove optimality, one would need to solve the infinite-dimensional continuous-time problem directly, which remains intractable in practice.

## IV. CASE STUDIES

Building on the iterative dispatch methodology, we extend the framework by incorporating dual analysis to build piecewise affine generation and price (LMP) trajectories. To demonstrate the price implications, two test systems are studied: an illustrative 5-bus system and a modified IEEE Reliability Test System 2019 (RTS-2019) [21]. These examples highlight the emergence of price discontinuities and the relationship between generation and price trajectories, providing insights into unit ramping behavior and its price implications. Furthermore, the results reveal price distortions and inefficient incentives that inherently arise in discrete-time formulations when viewed from a continuous-time perspective.

All optimizations are solved by CPLEX v12.7.1 in MATLAB on a ThinkPad X1 (2021) with an Intel(R) Core(TM) i5-1135G7 CPU. Parametric solutions are obtained using the publicly available MPT3 toolbox. The convergence criterion is set to a predefined tolerance of 0.001 MW. Continuous-time generation and price trajectories are expressed as piecewise functions of time, as detailed in [22].

### A. Results in the Illustrative 5-bus System

In the illustrative 5-bus system, physical parameters of units and continuous-time load profiles (seventh-degree polynomials) are taken from [20]. The iterative dispatch methodology is implemented by reorganizing the dispatch process and incorporating dual analysis. After three iterations, feasible continuous-time generation trajectories (Gen1 and Gen2) and price (LMP) trajectories are obtained, with 11 segment endpoints satisfying the convergence criterion. For clarity, numerical results are reported with two decimal places.

As shown in Fig. 3, the first iteration covers the full period [0, 60] minutes, and the subsequent iterations refine the solution over the updating ranges [9.62, 51.69] and [27.80, 31.65]. The final continuous-time price trajectories and their endpoints are also marked. Each iteration corresponds to a change in the critical region, during which a price jump occurs.

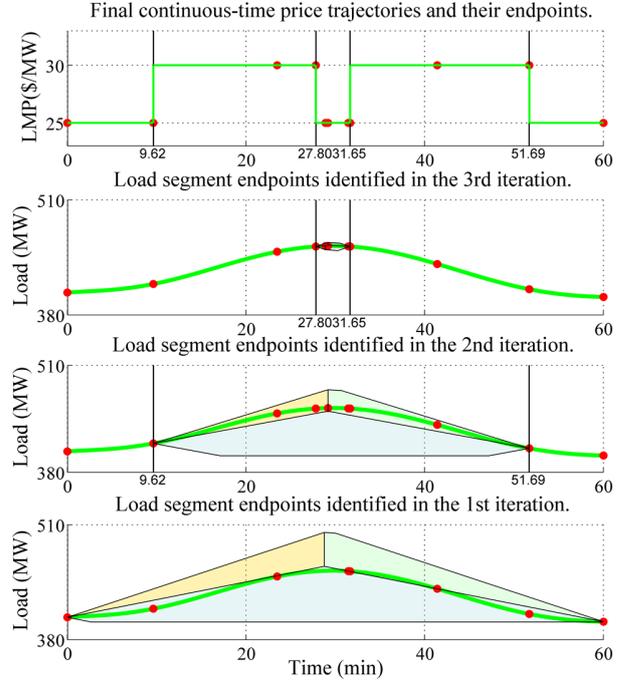

Fig. 3. Final continuous-time price trajectories and load segments obtained using the iterative dispatch methodology.

These price jumps are verified by identifying the instants of marginal-unit switching. To further illustrate this result, we combine generation and price trajectories to interpret how ramping capabilities affect price. In Fig. 4, the red line shows the final continuous-time generation trajectories $x^C(t)$, while the blue line represents the linear generation trajectories $x^D(t_n)$ obtained using the discrete-time adaptive dispatch. The red points mark the results at the interval endpoints.

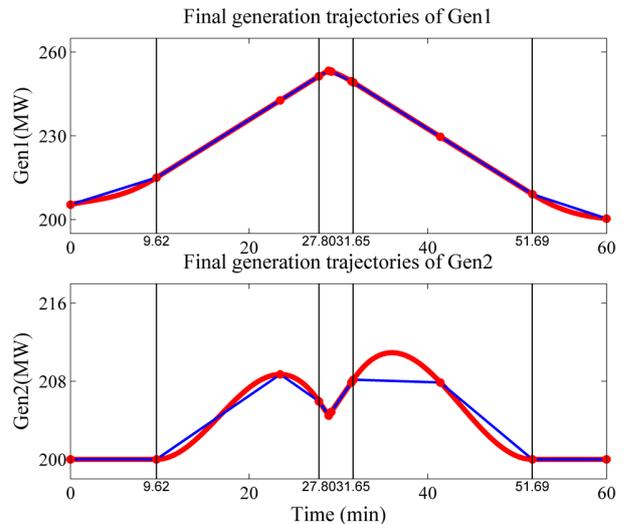

Fig. 4. Final continuous-time generation trajectories (red line) and linear trajectories from discrete-time adaptive dispatch (blue line).

As shown in Fig. 3, during the time span [0, 9.62], the continuous-time load increases gradually while Gen2 remains at its minimum output, making Gen1 the marginal unit. For an infinitesimal load perturbation $\delta D(t) \to 0$, Gen1's generation capacity and ramping capability are sufficient, and its marginal cost (25 \$/MW) therefore sets the market clearing price (LMP).

At $t_2$=9.62 min, the price jumps to 30 \$/MW and remains constant over [9.62, 27.80]. As shown in Fig. 4, throughout this time span [9.62, 27.80], Gen1 reaches its upward ramping limit. Although its generation capacity is sufficient, its ramping capability is exhausted. Hence, Gen1 can no longer serve as the marginal unit. Gen2 then becomes marginal, and the price is set at its bid of 30 \$/MW. This result is consistent with the corresponding piecewise affine representation. At the same time, Gen2 operates without binding unit-specific constraints, and its generation trajectories follow the same polynomial order as the load profiles. This confirms that the price over [9.62, 27.80] is determined by Gen2.

At $t_4$=27.80 min, both the generator and price trajectories undergo significant changes. As shown in Fig. 4, Gen1 and Gen2 are both ramping, but with different rates. Although the load continues to increase, Gen2 reaches its downward ramping limit, and Gen1 compensates by raising output due to its lower marginal cost. From this instant, Gen2's trajectory becomes linear, whereas Gen1 follows a seventh-order polynomial closely matching the load variation. Consequently, Gen1 becomes the marginal unit and sets the market price at 25 \$/MW.

During the subsequent parts of time span [27.80, 31.65], Gen2 gradually increases its output. As it reaches its upward ramping limit, Gen1 correspondingly reduces its output. Although Gen1's output decreases, it still provides the marginal response to infinitesimal load perturbation due to its lower marginal cost. At this instant, Gen2's trajectory is linear, whereas Gen1 follows a seventh-order polynomial that closely matches the load variation. Consequently, Gen1 remains the marginal unit, and the price stabilizes at 25 \$/MW.

This raises the question: why does Gen2 increase its output despite not being marginal? This ramping behavior may appear suboptimal from an instantaneous cost perspective; however, it arises from the requirement to satisfy future physical balance, since generation levels cannot change instantaneously. The ramping capability at any instant is constrained by the generation levels of preceding instants. This demonstrates the intertemporal coupling of units' ramping behavior—current generation adjustments are linked to future load requirements and, consequently, shape both the generation and price trajectories.

Similar analyses can be performed for other load segments. The final trajectories are determined once all trajectories are confirmed feasible and the price trajectories are verified to consistently align with the units' generation trajectories.

We now evaluate the price implications of the discrete-time model from a continuous-time perspective. A 5-min dispatch model is used to quantify the price difference between discrete-time and continuous-time models. Fig. 5 compares the 5-min LMP with the continuous-time LMP. The discrete-time price signal exhibits a noticeable time lag because it remains unchanged within each discrete-time dispatch interval.

This delay results in price distortions, leading to inefficient incentives. As shown in Fig. 5, the 5-min LMP reacts only at discrete-time dispatch intervals of 10, 30, 35, and 55 minutes. The shaded time spans of [9.62, 10], [27.80, 30], [31.65, 35], and [51.69, 55] minutes experience the time lag. During the shaded time spans, the 5-min model fails to capture intra-interval price jumps, thereby causing a loss of market efficiency.

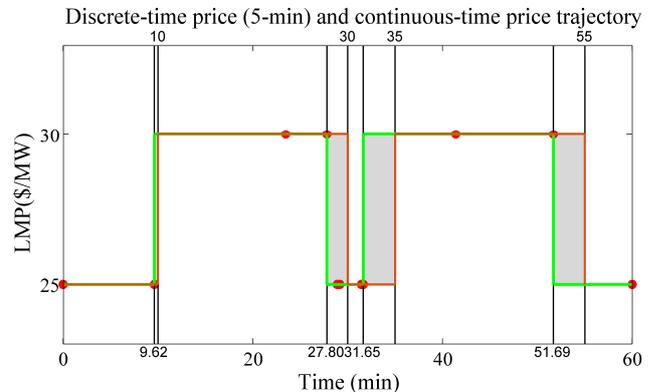

Fig. 5. Discrete-time 5-min LMP (orange line), continuous-time LMP (green line), and the price difference between them (shaded area).

When such delayed prices are applied in market settlement, Gen1 and Gen2—despite producing identical outputs—receive different revenues. This discrepancy in unit revenues illustrates the inefficient market incentives created by discrete-time price distortions. In Table I, the discrepancy in unit revenues between Gen1 and Gen2 is quantified. The price difference is defined as the continuous-time price minus the 5-min LMP. The discrepancy in unit revenues is calculated by multiplying this price difference by the continuous-time generation trajectories and integrating over each shaded time span. The discrepancy in unit revenues reflects distortions in market incentives: A positive value indicates excess revenue, whereas a negative value represents a revenue loss when viewed from a continuous-time perspective.

TABLE I
DISCREPANCIES IN UNIT REVENUES UNDER PRICE DISTORTIONS

| Shaded time span (min) | Gen1 (\$) | Gen2 (\$) |
| --- | --- | --- |
| [9.62, 10] | 6.82 | 6.33 |
| [27.80, 30] | -46.22 | -37.56 |
| [31.65, 35] | 68.63 | 58.57 |
| [51.69, 55] | -56.84 | -55.13 |

Here, we select the continuous-time generation trajectories for these comparisons; the discrepancies would be even larger if we employed the 5-min constant or linearly interpolated generation trajectories. These results highlight the model errors inherent in discrete-time formulations compared with the continuous-time formulation. As the time resolution increases, the shaded time spans—and the associated discrepancies in unit revenues—gradually decrease. In the limit of infinite resolution, the effect of such price jumps becomes manageable.

*B. Results in the Modified IEEE RTS-2019*

The modified IEEE RTS-2019 [21] is employed to validate the proposed iterative dispatch methodology. Generating units with very low or zero marginal costs, primarily renewables, are excluded. The remaining units with identical marginal cost coefficients are aggregated into representative units, resulting in 39 generators in the modified system. The generating units are indexed in ascending order of their marginal costs.

Continuous-time load profiles (Fig. 6) are constructed from 5-min real-time load data collected on January 1, 2020, and scaled to the hourly horizon using a 12th-degree polynomial fit.

After 67 iterations, the continuous-time load profiles are segmented by 80 endpoints. The corresponding generation and price trajectories are represented by piecewise affine functions. Detailed data and the final piecewise affine functions for these segments are provided in [22].

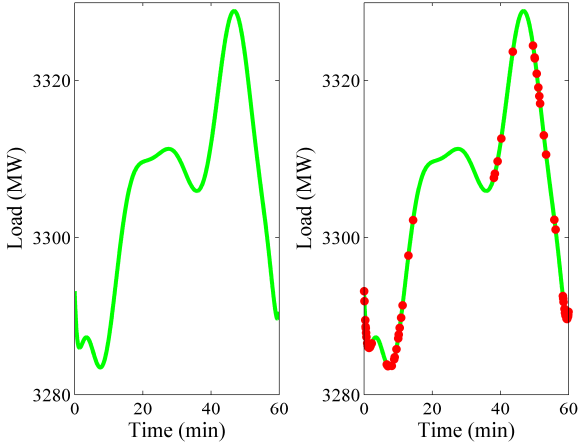

Fig. 6. Continuous-time load profiles for the IEEE RTS-2019. (a) Original profiles; (b) Final load segments with 80 endpoints.

With the generation and price trajectories, we analyze unit behavior and price in detail. Among the 39 units, 16 operate during this hour, while the remaining 23 remain offline. Among the 16 online generators, 8 operate at maximum output and thus cannot serve as marginal units (Gen1–Gen8), while the remaining 8 exhibit ramping behavior (Gen9–Gen16). The identification of marginal units is based on their marginal costs. The analytical objective is to determine the active marginal unit and verify whether the observed price trajectories are consistent with its behavior.

Ideally, if ramping capability were unlimited, Gen9–Gen15 would remain at maximum output, while Gen16 would serve as the marginal unit for the entire period [0, 60]. When ramping limitations are considered, however, the situation becomes more complex. Fig. 7 illustrates the ramping behaviors of the eight relevant units. In the early stage of the load drop, several units were required to provide ramping support. As the load evolved, Gen9–Gen15 quickly returned to maximum output to maintain overall economic efficiency, leading to frequent switches in the marginal unit. For example, Gen9's ramping behavior lasted only a few minutes—too short to be captured by discrete-time models but clearly visible in the continuous-time formulation. A similar result occurred near the end of the load drop, where the ramping behaviors of Gen13–Gen15 triggered further marginal-unit switching.

These ramping behaviors directly translate into fluctuations in the price trajectories. As shown in Fig. 8, the LMP exhibits frequent variations corresponding to switches in the marginal unit, particularly at the beginning of the period when a small price dip occurs. This dip arises because the marginal unit at that time is Gen8, the lowest-cost generator. As ramping scarcity gradually subsides, low-cost units resume operating at their maximum output, leading to a corresponding increase in the LMP.

From Fig. 7 and Fig. 8, it is evident that several units intentionally reduce their output in advance to preserve ramping capability for upcoming time spans. Such anticipatory adjustments ensure that sufficient flexibility is available to satisfy future ramping requirements. From a continuous-time perspective, this intertemporal coupling between ramping behaviors and price jumps can be captured. When a unit reduces output at the current instant to reserve ramping capability for future changes, the LMP at that instant reflects the marginal cost associated with switching the marginal unit. Accordingly, when an infinitesimal load perturbation $\delta D(t)$ occurs, it is absorbed by the pre-arranged ramping capability associated with the unit offering the lowest marginal cost.

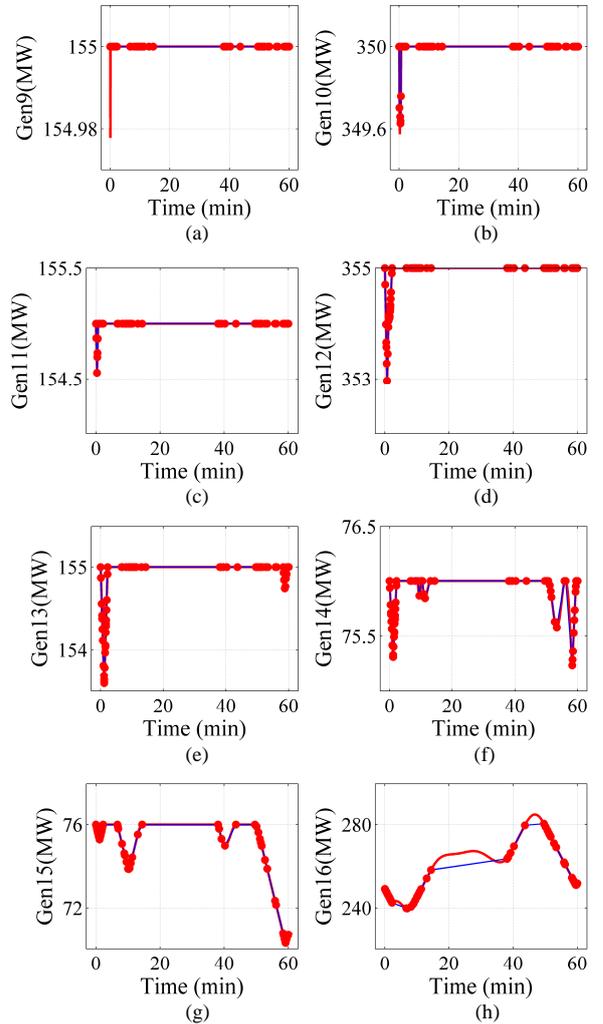

Fig. 7. Continuous-time generation trajectories: (a) Gen9; (b) Gen10; (c) Gen11; (d) Gen12; (e) Gen13; (f) Gen14; (g) Gen15; (h) Gen16.

In summary, ramping capability serves a similar role to generation capacity in determining the marginal unit at each instant. The Lagrange multiplier of the system-wide power balance constraint represents both the value of an infinitesimal load perturbation and the marginal cost of the marginal unit. Price jumps occur precisely when marginal units switch, reflecting natural market outcomes rather than distortions.

Furthermore, the intertemporal coupling of ramping behaviors directly shapes continuous-time price trajectories, where the LMP at any instant reflects both current system conditions and anticipated future requirements. From this continuous-time perspective, discrete-time dispatch models inherently distort prices because they fail to capture intra-temporal variations, leading to inefficient and misleading market incentives.

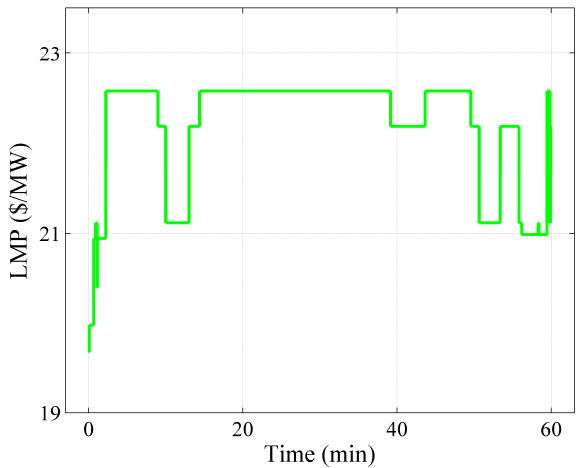

Fig. 8. Final continuous-time price trajectories (LMP) for the IEEE RTS-2019.

## V. Conclusion and Future Work

This paper develops a dual analysis framework for continuous-time economic dispatch, enhancing the understanding of discrete-time formulations in capturing intra-temporal variations and the resulting price distortions. The iterative dispatch methodology is extended to construct dual price trajectories, enabling consistent primal-dual analysis from a continuous-time perspective. Results show that price signals—represented by the Lagrange multipliers of system-wide balance constraints—evolve continuously except at critical instants, where discontinuities naturally arise from marginal-unit switching.

The analysis further clarifies that ramping constraints introduce explicit intertemporal price coupling that links adjacent time intervals. As a result, the principal value of the continuous-time model lies not in reproducing static scarcity effects, but in revealing how ramping constraints generate intertemporal price interactions and price discontinuities that are obscured in discrete-time models.

Case studies on an illustrative 5-bus system and the modified IEEE RTS-2019 validate the proposed methodology, demonstrating how ramping constraints and intertemporal coupling shape continuous-time price trajectories. The analysis further reveals that marginal-unit transitions determine the instant of price jumps, with instantaneous prices reflecting not only current operating conditions but also anticipated future requirements. Analyzing generation and price trajectories is therefore essential to understand the misleading incentives inherent in discrete-time formulations.

The continuous-time model primarily serves as an analytical framework for identifying and quantifying price distortions arising from the improper treatment of intertemporal coupling. Future work will focus on enhancing its practical applicability by explicitly incorporating unit commitment, transmission constraints, and stochastic factors. These extensions will offer deeper insights into system dynamics and price formation mechanisms in real-world electricity markets.